%% file: FHM.tex
\documentclass[a4paper,11pt]{article}

\usepackage{jinstpub} 

\usepackage{xspace}
\usepackage{caption}
\usepackage{subfigure}
\usepackage{xcolor}
\usepackage{color}
\usepackage[normalem]{ulem}
\usepackage{bm}

\usepackage{lineno}

\input{commands}

\title{\boldmath First results on FHM – a Floating Hole Multiplier}

\author[a,1]{V. Chepel,\note{Corresponding author.}}
\author[b,c]{G. Martinez-Lema,}
\author[b,c]{A. Roy,}
\author[b]{and A. Breskin}

\affiliation[a]{LIP-Coimbra and Department of Physics, University of Coimbra, 3004-516 Coimbra, Portugal}
\affiliation[b]{Dept. of Astrophysics and Particle Physics, Weizmann Institute of Science, Rehovot, Israel}
\affiliation[c]{Unit of Nuclear Engineering, Ben-Gurion University of the Negev, Beer-Sheva, Israel}

\emailAdd{vitaly@coimbra.lip.pt}

\abstract{\input{src/abstract}}

\keywords{
\\ Noble liquid detectors
\\ Micropattern gaseous detectors
\\ Charge transport, multiplication and electroluminescence in rare gases and liquids
\\ Time Projection Chambers
\\ Dark Matter detectors (WIMPs, axions, etc.)

}

\begin{document}
\maketitle
\flushbottom

\input{src/introduction}
\input{src/setup}

\input{src/results}

\input{src/discussion}

\input{src/conclusions}

\appendix

\acknowledgments
\input{src/acknowledgments}

\bibliographystyle{JHEP}
\bibliography{FHM}


\end{document}

%% file: commands.tex
\newcommand{\dvthgem}{\ensuremath{\Delta V_{\textrm{THGEM}}}\xspace}
\newcommand{\dvdrift}{\ensuremath{\Delta V_{\textrm{drift}}}\xspace}
\newcommand{\dvextr}{\ensuremath{\Delta V_{\textrm{extr}}}\xspace}

\newcommand{\sone}{\ensuremath{S1}\xspace}
\newcommand{\stwo}{\ensuremath{S2}\xspace}

\newcommand{\ignoreblock}[1]{}

%% file: src/abstract.tex

A proof of principle of a novel concept for event recording in dual-phase liquid xenon detectors --- the Floating Hole Multiplier (FHM) --- is presented. It is shown that a standard Thick Gaseous Electron Multiplier (THGEM), freely floating on the liquid xenon surface permits extraction of electrons from the liquid to the gas. Secondary scintillation induced by the extracted electrons in the THGEM holes as well as in the uniform field above it was observed. The first results with the FHM indicate that the concept of floating electrodes may offer new prospects for large-scale dual-phase detectors, for dark matter searches in particular.

%% file: src/introduction.tex
\section{Introduction}
\label{sec:intro}

Two-phase noble-liquid detectors play an important role in modern particle and astroparticle experiments requiring low energy threshold, specific event signature and extremely low radioactive background. Among them, the search for dark matter in the form of WIMPs (Weakly Interacting Massive Particles) and more recently neutrino physics are the most notable applications. More complete information and references on this subject can be found in the following reviews and books~\cite{Akimov:2021book, Bolozdynya:2010book,  Aprile:2008book, Chepel:2013, AprileDoke:2010,  Aalbers:2022whitepaper}.

The two most outstanding properties of liquefied noble gases from the point of view of particle detection are: i) the possibility to measure both scintillation and ionisation signals, and ii) the possibility of electron extraction from liquid to gas. The first provides particle identification as the electron-ion recombination along the track depends on the excitation and ionisation density induced by a particle in the liquid (which in turn depends on the particle and its energy). On the other hand, extraction of the electrons to the gas allows signal amplification similar to gaseous detectors. The high purity of the medium --- essential for collecting  the electrons from large distances --- makes charge multiplication somewhat difficult due to a strong VUV photon feedback. However, gas electroluminescence facilitates the reliable detection of individual electrons extracted from liquid xenon (LXe), with a high signal-to-noise ratio, as has been reported by several authors, both in laboratory and large-scale experiments (e.g.~\cite{Chepel:2013, Akimov:2021book, Aalbers:2022whitepaper}). In these two-phase detectors, it is customary to deploy a wire grid a few mm above the liquid surface so that the electroluminescence develops in a nearly uniform field between the surface and the grid. 

The presence of the liquid--gas interface in a detector has a number of drawbacks. Besides the technical challenges that are growing fast with detector upscaling, some undesirable phenomena occur at the surface. Spontaneous electron emission has been observed in liquid xenon detectors, the nature of which is still not understood (see, for example,~\cite{Akimov:2016single, Edwards:2007single, Burenkov:2009, Santos:2011single, Aprile:2013single, Sorensen:2017single, Bodnia:2021single}). Part of it may be due to electrons trapped under the liquid surface that may drift or diffuse in horizontal directions and emerge at some distance from their original location at the liquid-gas interface~\cite{Bolozdynya:1991undersurface, Akimov:2016single}. Surface micro-instabilities and ripples (for example, those induced by gas circulation) may also disturb the electron emission process.

Therefore, alternatives to the traditional design consisting of two multiwire grids --- one just below the liquid surface and the other a few mm above it --- are being considered by several groups including both dual- and single-phase (liquid) configurations. We refer the reader to a recent article~\cite{Breskin:2022novel} for a complete list of references and new ideas on signal recording in noble-liquid detectors. 

\begin{figure}[h]
\centering
\includegraphics[scale=0.3]{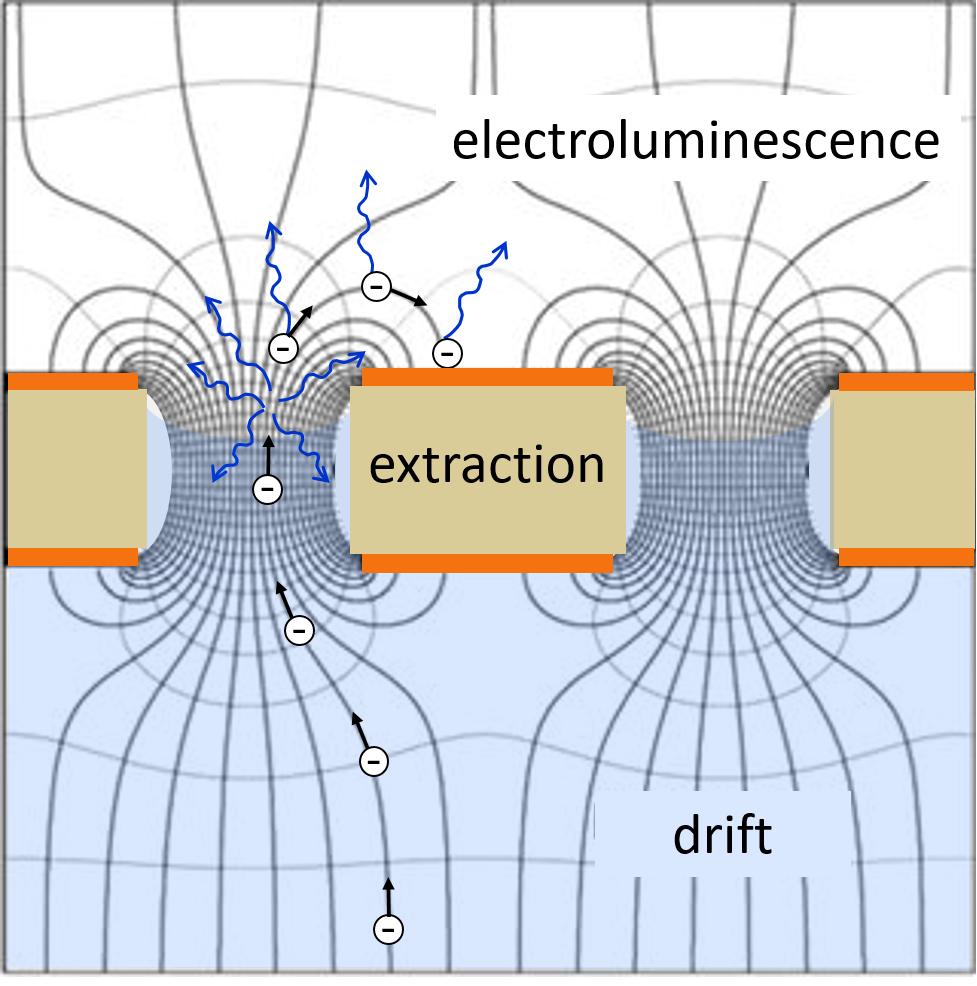} 
\caption{The concept of the floating electrode using a THGEM. Electrons from the liquid are collected into the THGEM holes, cross the liquid surface in the hole and are extracted into the gas where they induce electroluminescence. The extraction field is defined by the voltage across the THGEM; the drift field in the bulk of the liquid --- by the potential difference between the cathode and the bottom face of the THGEM; the field above the plate is a superposition of the field inside the hole and its vicinity and a uniform field that may be set using an additional electrode in the gas phase. This field can be strong enough to extract most of the electrons from the hole and induce electroluminescence also in the uniform field above the THGEM.}
\label{fig:principle}
\end{figure}

The motivation behind this work is to propose a novel concept for the dual-phase configuration, with substantially reduced surface-related instabilities. The idea, illustrated by Fig.~\ref{fig:principle}, consists in using perforated electrodes freely floating on the liquid surface. Different configurations are possible, but a THGEM (THick Gaseous Electron Multiplier~\cite{Breskin:2009THGEM, Sauli:2020book}) seems the most natural solution for LXe, given a large difference between the density of the liquid (about 2.9~g/cm$^3$) and that of the FR4-made (1.85~g/cm$^3$) THGEM substrate. In this configuration, with most of the liquid surface in contact with the bottom face of the THGEM, only a small part of the liquid facing the THGEM holes is exposed to the gas. This results in a substantial reduction of the undesirable free-surface related effects, although, on the other hand, the electron extraction efficiency into gas may be questioned due to the configuration of the field lines \cite{Erdal:2018}. Furthermore, the bottom face of the perforated electrode may be coated with a VUV sensitive photocathode (e.g. CsI), to permit also the detection of photons emitted in primary scintillation (see, for example, discussion in~\cite{Breskin:2022novel}). 

In this article, we describe the first experimental results obtained with a THGEM electrode floating on the surface of LXe, thus validating the concept.

\vspace{0.5cm}

%% file: src/setup.tex
\section{Experimental setup}
\label{sec:setup}
\begin{figure}[h]
\centering
\includegraphics[scale=0.3]{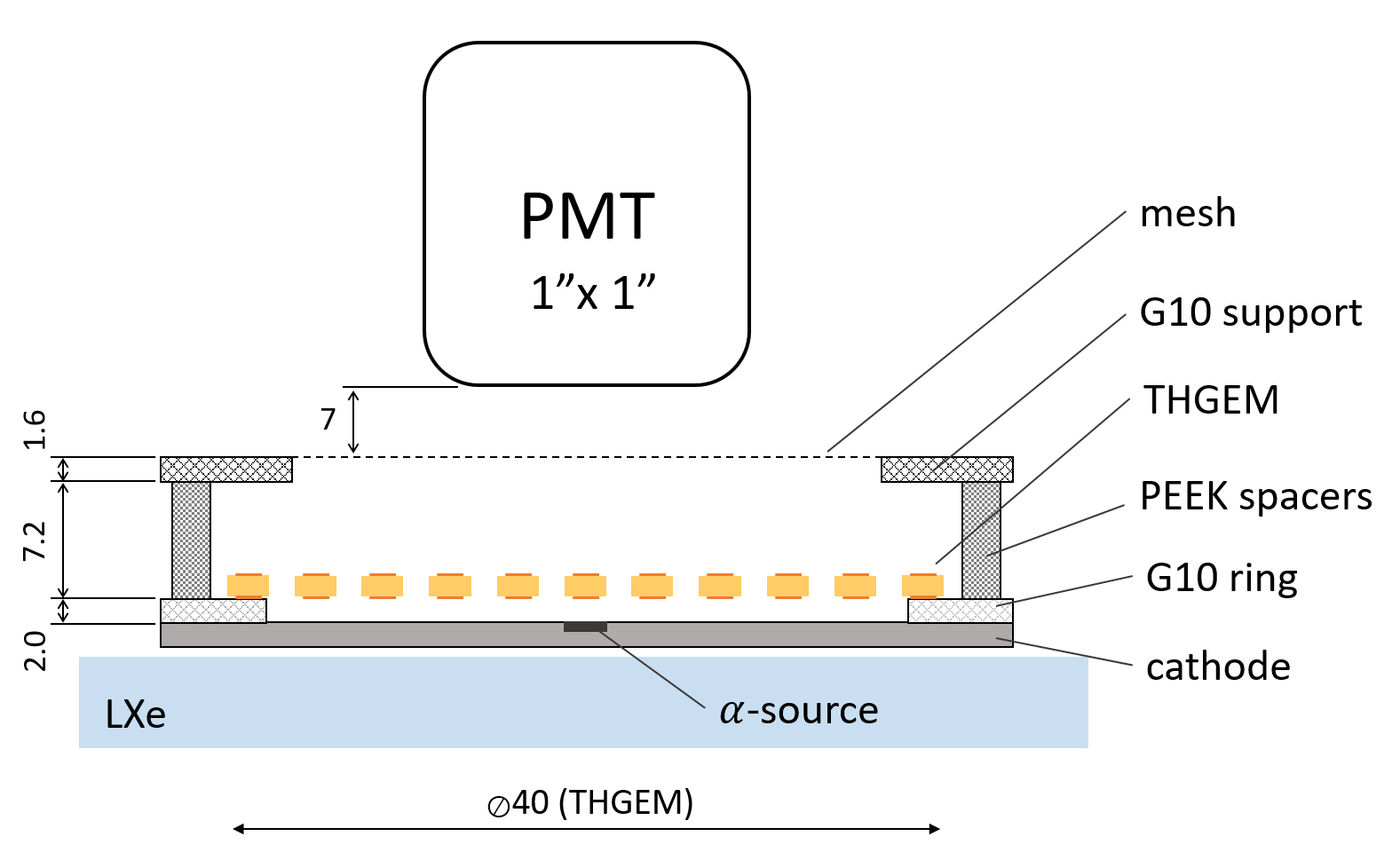} 
\hfill
\includegraphics[scale=0.3]{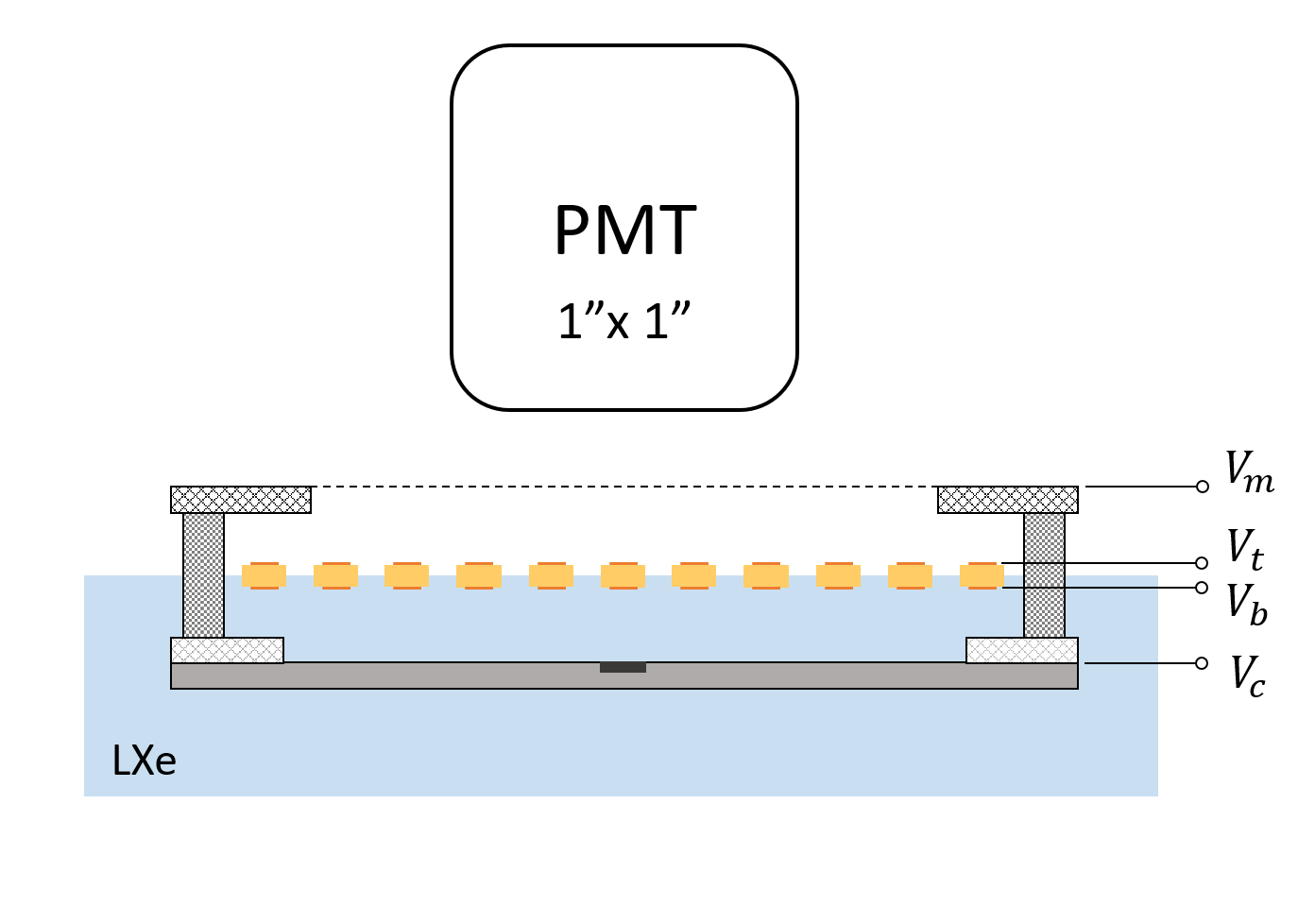}
\caption{Schematics of the experimental setup. Left: THGEM electrode in its initial position with LXe below the cathode. Right: The THGEM electrode floating on the LXe surface.}
\label{fig:setup}
\end{figure}

The experimental setup is schematically shown in Fig.~\ref{fig:setup}. The electrode system is constituted by a stainless steel cathode of 55~mm diameter at the bottom, a THGEM that is allowed to move freely in vertical and horizontal directions, a stainless steel mesh above the THGEM and a VUV sensitive photomultiplier. A small (6~mm diameter)  $^{241}$Am alpha-source is placed at the centre of the cathode. The source activity is about 40~Bq. 
\vspace{0.7cm}

\begin{figure}[h]
\centering
\includegraphics[scale=0.36]{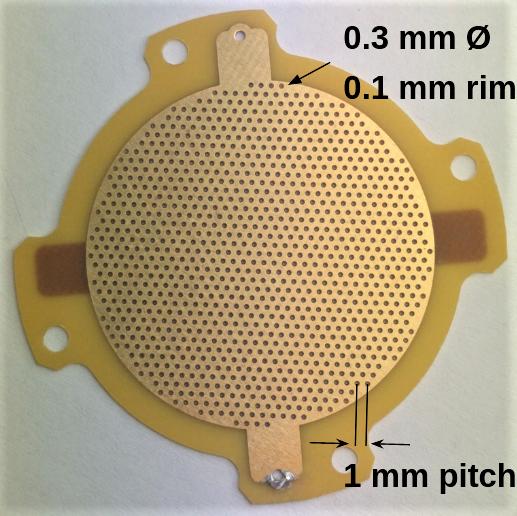} 
\hspace{2cm}
\includegraphics[scale=0.06]{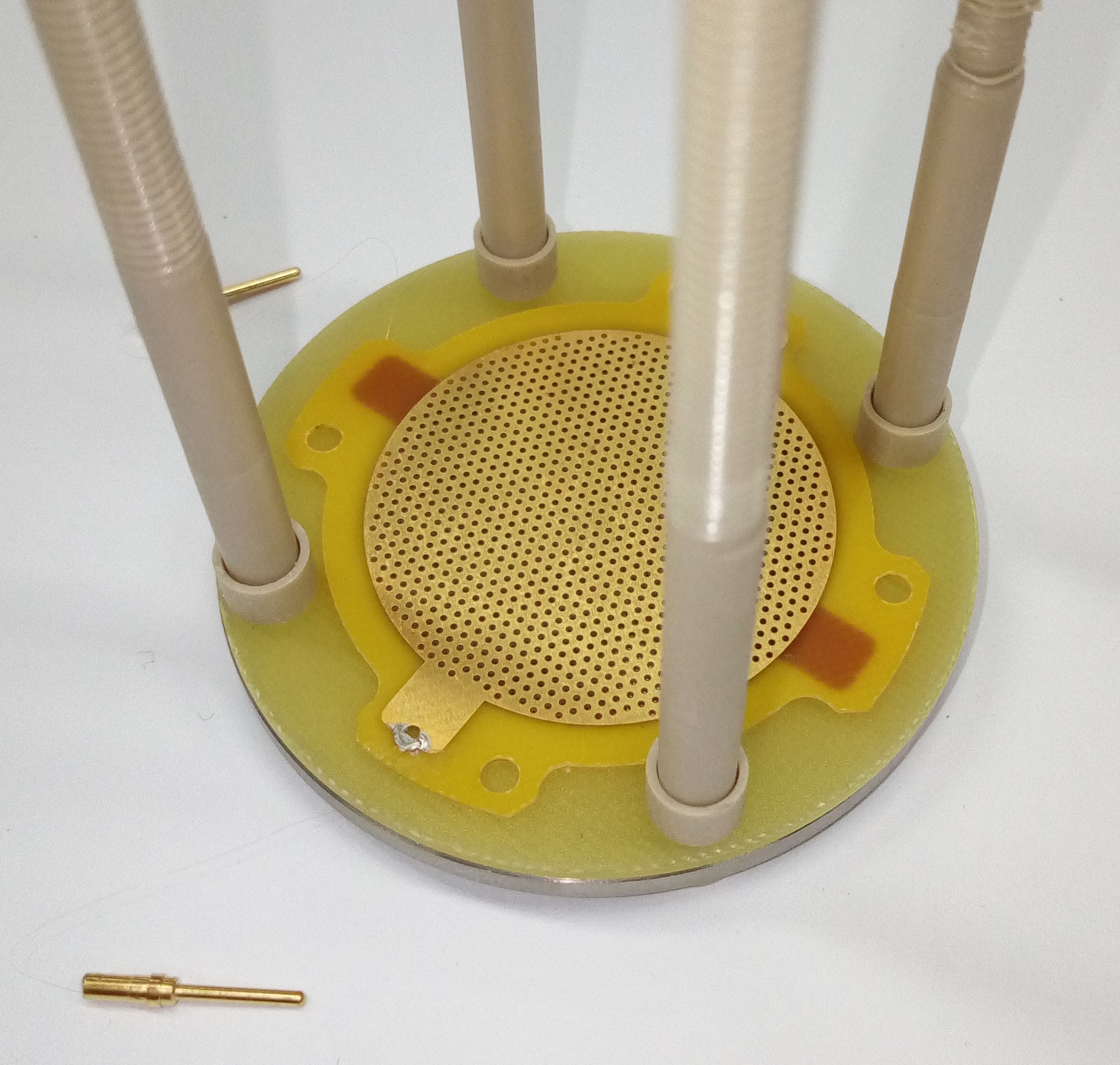}
\caption{Left: Details of the 40~mm diameter 0.45~mm thick THGEM electrode used in this setup. The hole diameter is~0.3 mm with a 0.1~mm rim. The holes are 1 mm apart. Right: Details of THGEM mounting between the insulating rods. The two contacts visible in the image are connected with 20~$\mu$m wires (invisible) to each side of the THGEM electrode.}
\label{fig:details}
\end{figure}

A THGEM electrode made of 0.45~mm thick FR4 with gold-plated copper cladding on both sides was used in this experiment (Fig.~\ref{fig:details}). The electrode is 40~mm in diameter with an active area of 34~mm. The holes are drilled in a hexagonal pattern, the hole  diameter being 0.3~mm, pitch of 1~mm and the etched rim around each hole is 0.1~mm wide. 

\vspace{0.7cm}

Details of the THGEM assembly are shown in Fig.~\ref{fig:details} (right). The whole assembly is kept together with 4 peek rods placed at such distances that the THGEM  can move freely between them, both in the horizontal and vertical directions. The electrical contacts to both sides of the THGEM electrode are made with 20~$\mu$m wires about 4~cm long to allow the unrestricted movement of the THGEM between the cathode and the mesh and also between the 4 peek rods. A transition to thicker wires is made with spring contacts fixed on an auxiliary G10 disc above the electrodes (not shown) to which the whole assembly is attached. 

The upper electrode is a woven stainless steel mesh with a pitch of about 0.6 mm,  soldered on top of a 1.6~mm thick G10 support. The mesh open diameter is 25 mm. On top of the cathode disc, another 2~mm thick G10 ring is placed to avoid contact of the THGEM electrode with the cathode when the system is in gas or in vacuum. Peek spacers are used to keep a fixed distance between the mesh and the cathode.

The dimensions are provided in Fig.~\ref{fig:setup}; the total distance between the cathode and the mesh is 10.8~mm. Before liquefaction, the THGEM is positioned on top of the 2~mm thick G10 ring covering the cathode. The length of the vertical path available for the THGEM movement is defined by the length of the peek spacers (7.0~mm) between the top surface of the G10 ring at the bottom and the lower surface of the G10 mesh support at the top (see Fig.~\ref{fig:setup}). In the horizontal direction, there is a space of about 2~mm from each side of the THGEM to the peek rods with spacers (see Fig.~\ref{fig:details}). The assembly also allows the THGEM to rotate in the horizontal plane within $\sim$30$^\circ$.

The photomultiplier tube (PMT) used in these measurements is a 1$\times$1-inch Hamamatsu PMT  with square quartz window, metallic body, bialkali photocathode and mesh-type dynodes, model R8520-406~\cite{Hamamatsu}. The PMT is placed in xenon gas at a distance of about 7~mm from the mesh electrode (see Fig.~\ref{fig:setup}). It was operated at a negative bias of --750~V (unless otherwise is stated); the anode signals were recorded with a digital oscilloscope (Tektronix MSO5204B).

During the experiments, one of the electrodes of the THGEM --- either bottom or top --- was kept at ground potential, the cathode was negatively biased, the potential on the mesh was normally kept positive, although a reversed bias was applied in some measurements. 

The whole assembly was inserted into the MiniX cryostat, described in detail elsewhere~\cite{Erdal:2019Thesis}. The xenon pressure was kept at 1.23~bar during the measurements, corresponding to a temperature of 170K. Before starting xenon condensation, the chamber was pumped and xenon gas circulated through it via a SAES hot getter, model PS3-MT3-R-2, at the rate of $\sim$1~slpm for about 12~hours. 

%% file: src/results.tex
\section{Results}
\label{sec:results}

PMT signals have been measured at different voltages applied to the cathode, THGEM electrodes and the mesh (see Fig.\ref{fig:setup}, right). We will use the following notations through this article:  \dvdrift~$=V_b-V_c$ for potential difference between the THGEM bottom face and the cathode, \dvthgem~$=V_t-V_b$ for that across the THGEM, and \dvextr~$=V_m-V_t$ for potential difference between the mesh and the top THGEM electrode.

\begin{figure}
    \centering
    \includegraphics[width=12cm]{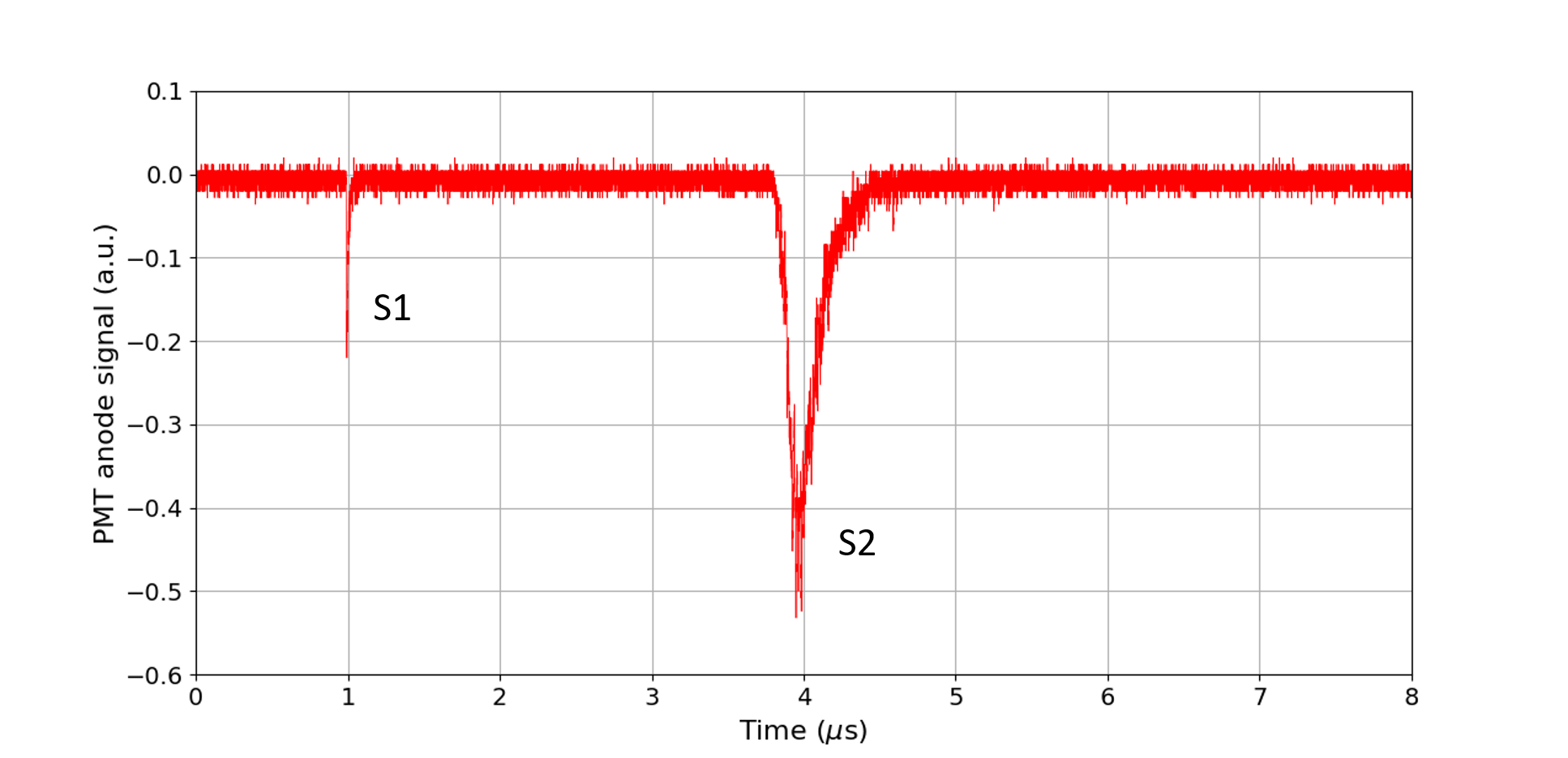}
    \caption{An example of a PMT waveform showing both \sone and \stwo pulses, measured at ${\dvdrift=400}$~V, \dvthgem$=2500$~V, \dvextr~$=0$~V. The chamber is partially filled with liquid xenon.}
    \label{fig:waveform_example}
\end{figure}

Fig.~\ref{fig:waveform_example} shows an example of a waveform measured at the PMT anode in the configuration of Fig.~\ref{fig:setup} (right) with \dvdrift $=400$~V, \dvthgem$=2500$~V and \dvextr~$=0$~V. The liquid level was estimated to be about 6.5~mm above the cathode. The oscilloscope was triggered by the primary scintillation induced by the alpha-particles (\sone), visible in the plot as a short pulse at $t=1~\mu$s. A second pulse is clearly visible after a few $\mu$s being attributed to secondary scintillation (\stwo) in xenon gas within the THGEM holes and/or their vicinity. 

\begin{figure}
    \centering
    \includegraphics[width=12cm]{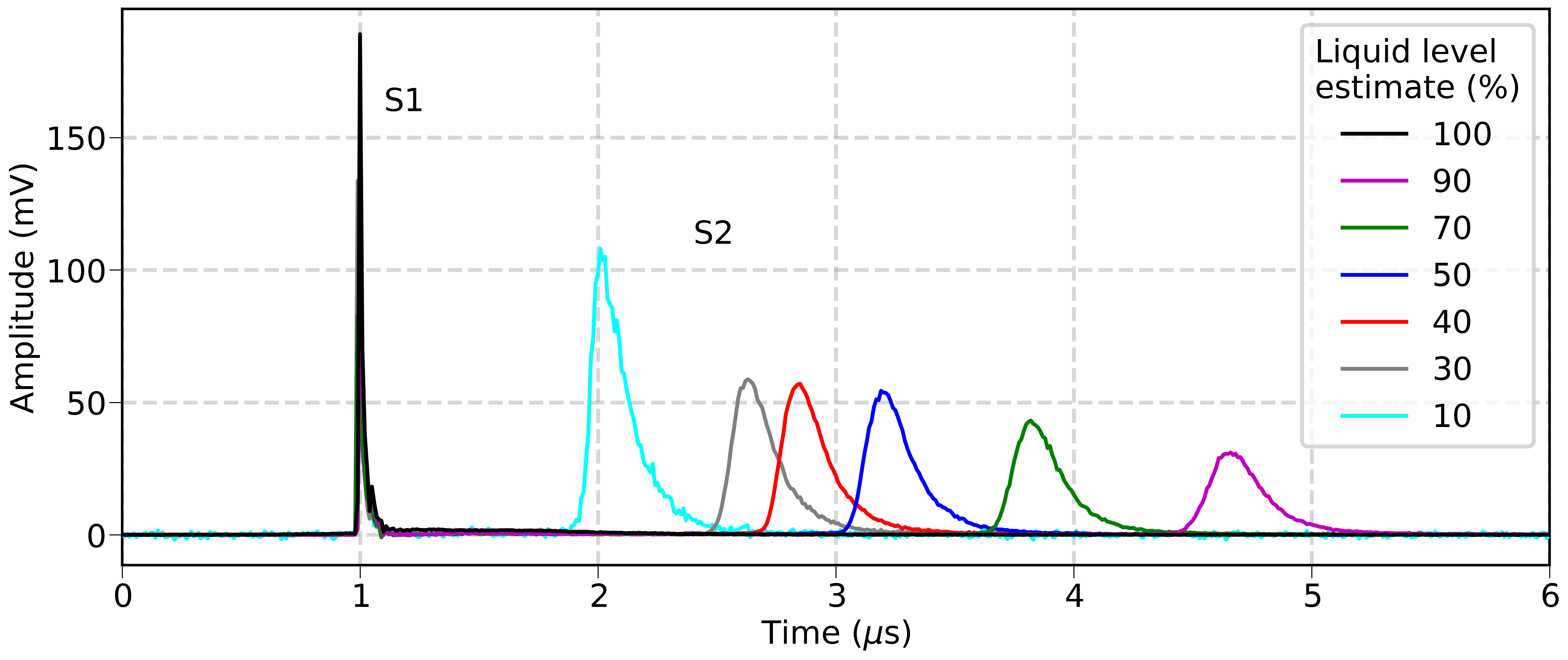} 
    \caption{Averaged PMT signal waveforms taken with different liquid levels. All the voltages are kept constant: \dvdrift $=400$~V, \dvthgem$=2500$~V, \dvextr~$=0$~V. The time difference between the \sone and \stwo signals indicates that the distance between the cathode and the THGEM increases --- clear evidence that the electrode is floating.  }
    \label{fig:waveforms_during_filling}
\end{figure}

The LXe level in the chamber was also varied during the measurements by adding some xenon or by recondensing it into a storage cylinder kept at liquid nitrogen temperature. The liquid level was not directly measured but could be inferred from the PMT signal shape,  namely from the time interval between \sone and \stwo, which reflects the electron drift time in the liquid. Waveforms measured at different levels of liquid are shown in Fig.~\ref{fig:waveforms_during_filling}. One could also observe the situation of the cathode being in gas and that of the liquid totally covering the THGEM (lack of \stwo) when the liquid level was above the upper limit for the THGEM vertical position (G10 mesh support --- see Fig.~\ref{fig:setup}). The waveforms here are averages of 1000 signals measured with a digital oscilloscope triggered on \sone. In this experiment, all voltages were kept constant, \dvdrift = 400~V, \dvthgem = 2500~V and ${\dvextr = 0}$~V. The liquid level was estimated from the number and duration of the cryopumping cycles (i.e. condensation of xenon vapours from the chamber to a storage cylinder). The precision of this procedure can be estimated as $\sim$0.5~mm.

\begin{figure}
    \centering
    \includegraphics[width=6cm]{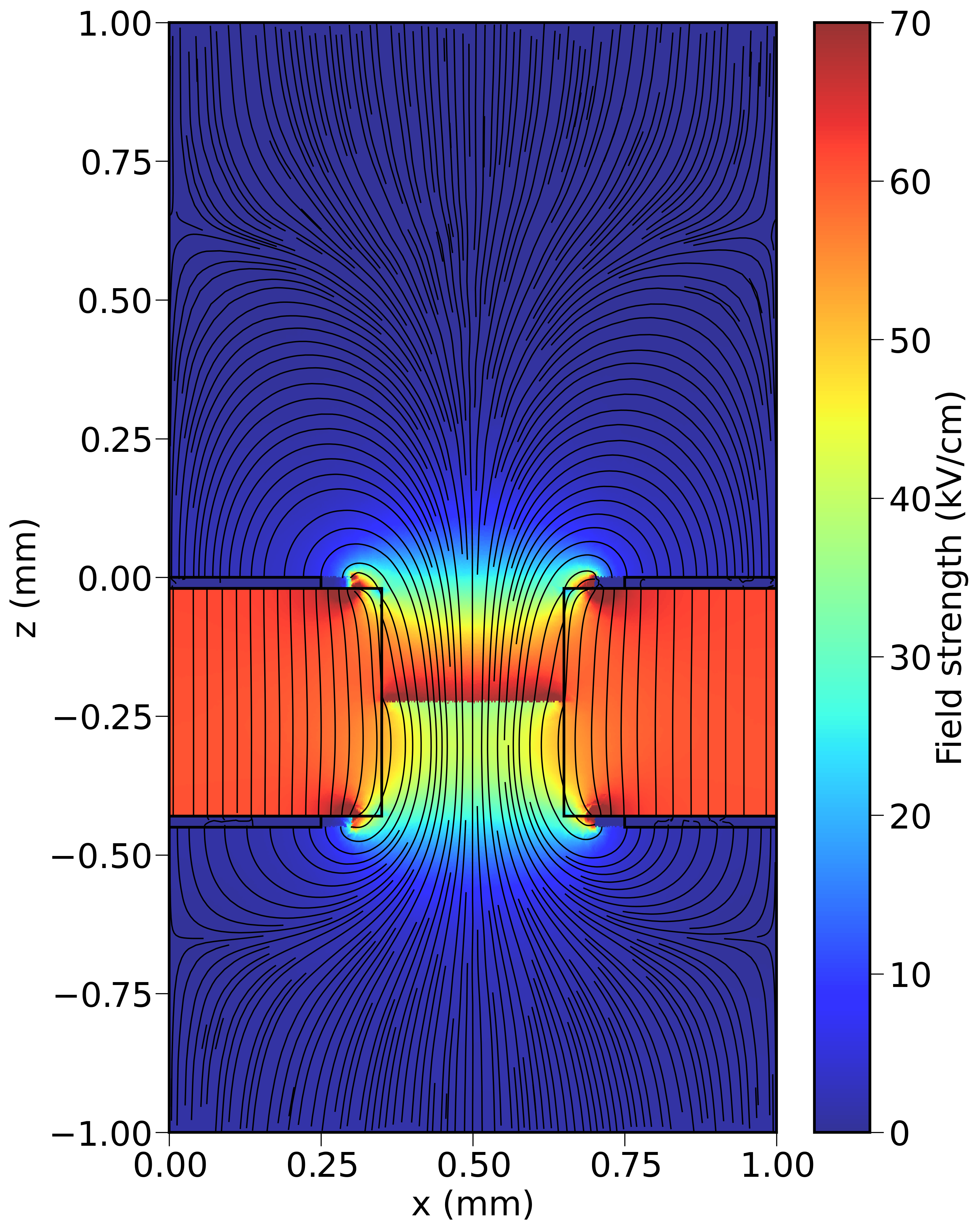} 
    \caption{Electric field strength and field lines in the vicinity of the THGEM holes simulated in COMSOL\copyright. A flat liquid-gas interface equidistant to both faces of the 0.45~mm THGEM is assumed. The voltage configuration is \dvdrift $= 400$~V, \dvthgem $= 2500$~V and \dvextr $= 0$. The drift gap is 4.0~mm and the extraction gap is 8.0~mm. }
    \label{fig:comsol}
\end{figure}

With zero field set above the THGEM during the course of the measurements, the second pulse is explained by secondary scintillation within the THGEM holes, partially in gas, or at their vicinity above the electrode (secondary scintillation in the liquid would require much higher field of $\sim$400~kV/cm, according to~\cite{Aprile:2014ELthreshold}). The calculated electric field in the vicinity of a THGEM hole, partially filled with liquid xenon, is shown in  Fig.~\ref{fig:comsol}.
Therefore, electrons are indeed extracted from liquid to gas (part of them, at least). The dependence of the time interval between \sone and \stwo on the liquid level clearly indicates that the THGEM is floating on the surface of the liquid (note, that as \dvdrift~ $=$~const, two factors contribute to the increase of the drift time with the liquid layer thickness --- the increase of the drift distance and the decrease of the electron drift velocity; both factors act in the same direction).

\begin{figure}
    \centering
    \includegraphics[width=7cm]{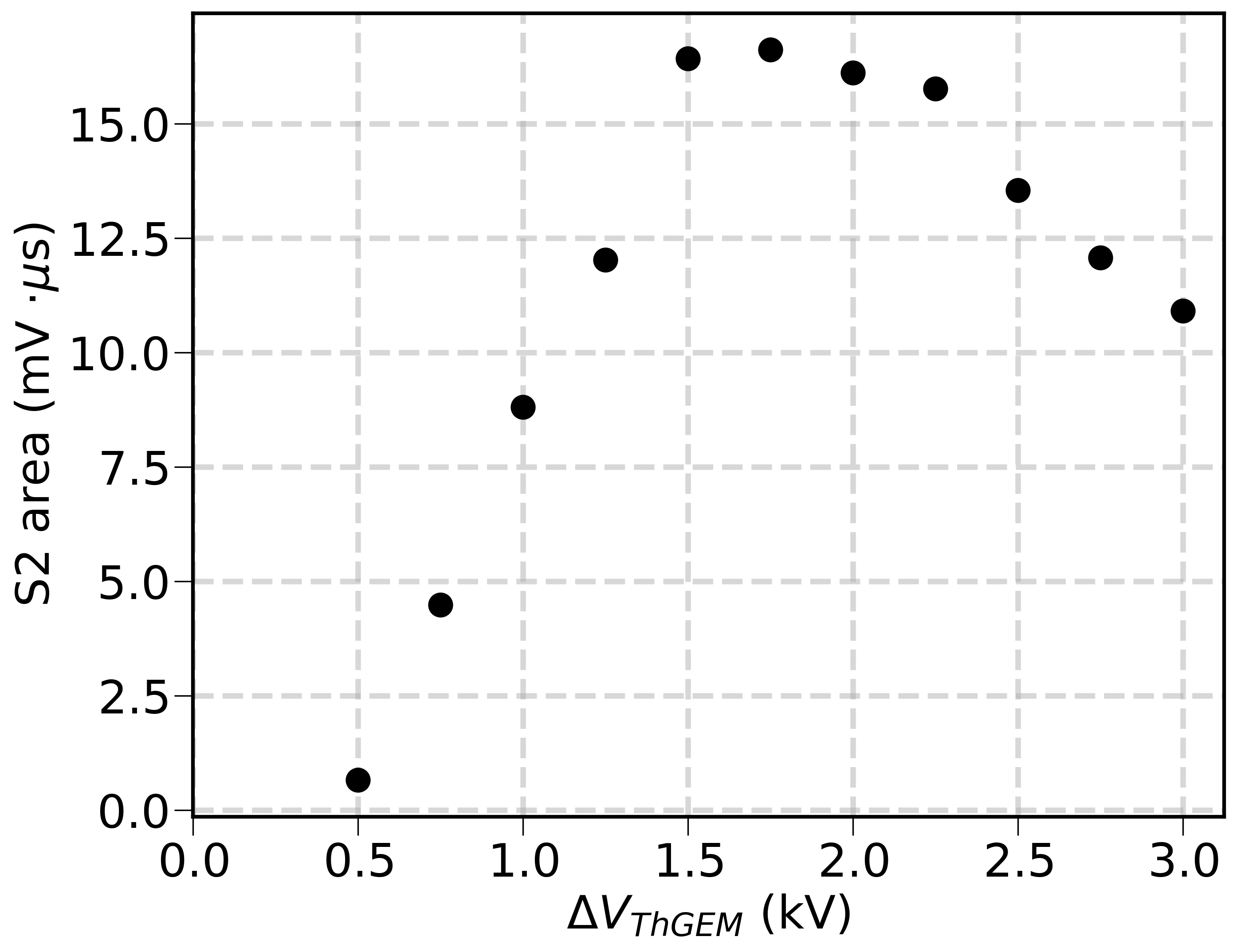} 
    \caption{Variation of the area of \stwo signals as a function of the voltage across the THGEM at fixed \dvdrift$=400$~V, \dvextr$=-500$~V (reversed field) and $V_{\textrm{PMT}} = -700$ V.
    }
    \label{fig:s2_area_vs_vthgem}
\end{figure}

The effect of the voltage across the THGEM on the light yield is shown in Fig.~\ref{fig:s2_area_vs_vthgem}. The initial linear rise is expected, however the saturation at $\dvthgem\approx$~1500~V and further decrease require an explanation. This might be due to the effect of the electric field in the THGEM holes on the liquid surface --- a dielectric in a non-uniform electric field is subject to a force in the direction of the field gradient. Xenon atoms are known to have high polarizability, so that the level of liquid xenon should rise in the holes as the voltage on the THGEM increases, thus reducing the size of the scintillation region in gas. For example, expansion of liquid argon into a bubble of argon gas in the THGEM holes was observed under an electric field as reported in~\cite{Tesi:2021}.

\begin{figure}
    \centering
    \includegraphics[width=12cm]{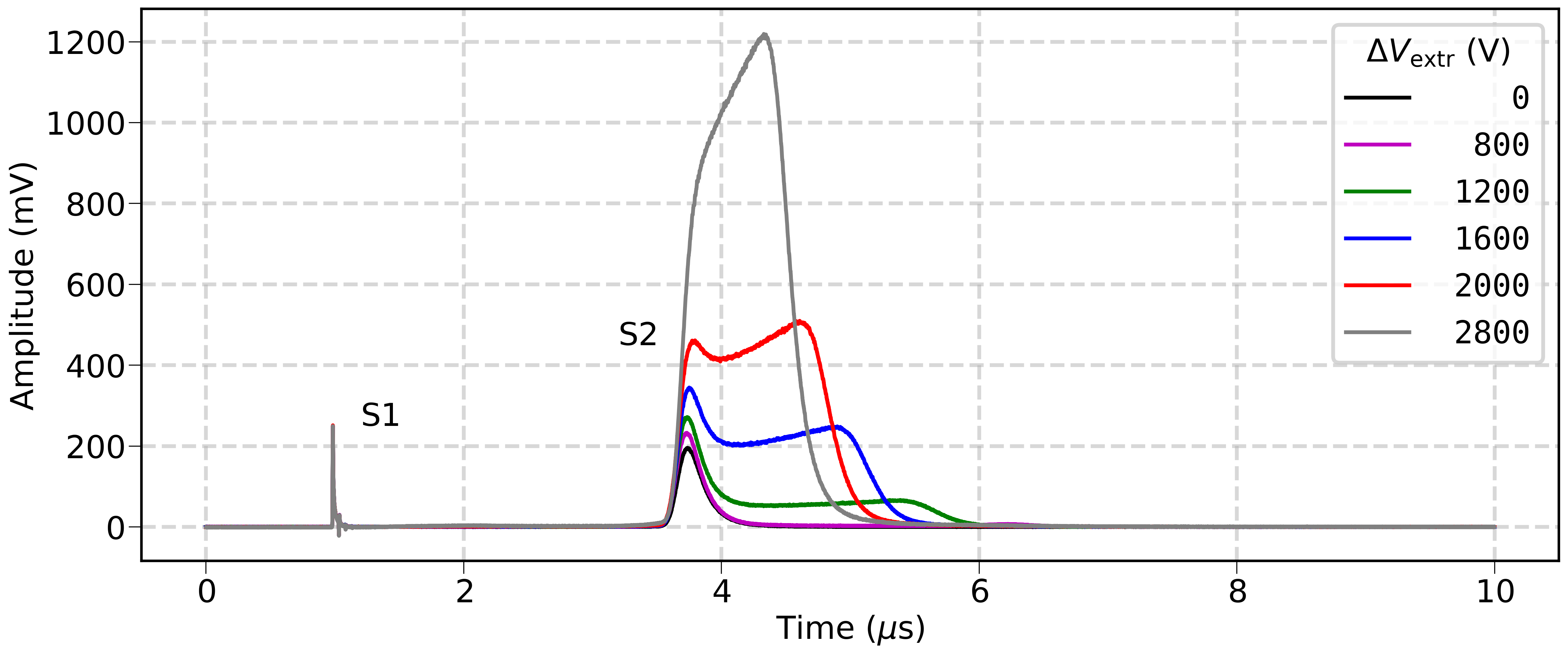} 
    \caption{Shape of \stwo signals for different values of extraction field, at fixed \dvdrift $=400$~V and \dvthgem $=2500$~V. }
    \label{fig:s2_shape}
\end{figure}

Applying some voltage between the THGEM top surface and the mesh, \dvextr, changes the \stwo shape (Fig.\ref{fig:s2_shape}). A long tail appears in the \stwo pulse that is explained by secondary scintillation induced by the electrons drifting in the region of uniform field between the THGEM and the mesh. The \stwo area gradually increases as \dvextr increases, while the tail becomes shorter since the electron drift velocity in gas increases with the field. At \dvextr$\approx1000$~V, the light yield in the uniform field is similar to that in the THGEM hole vicinity. 

\begin{figure}
    \centering
    \includegraphics[width=7cm]{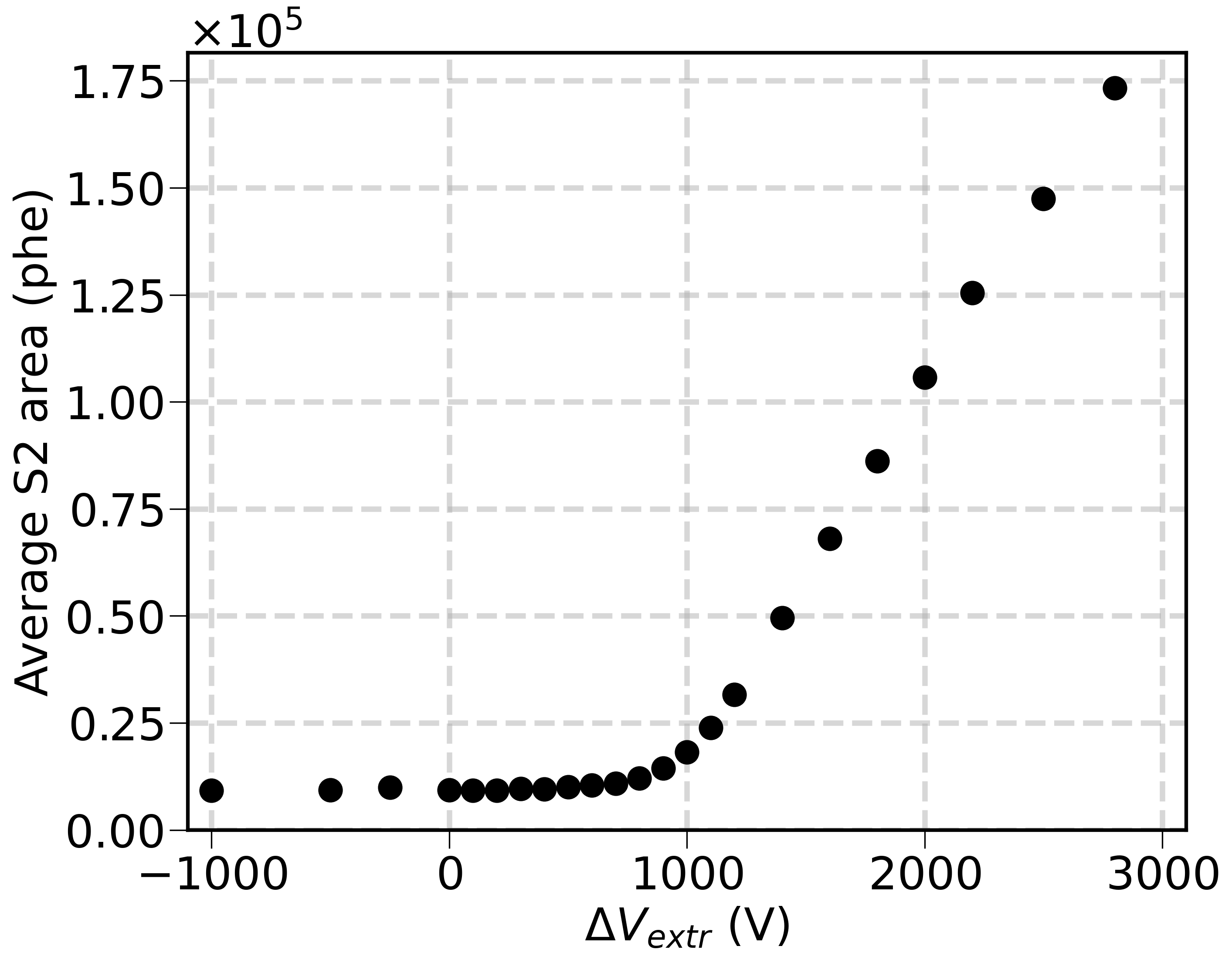} 
    \includegraphics[width=7cm]{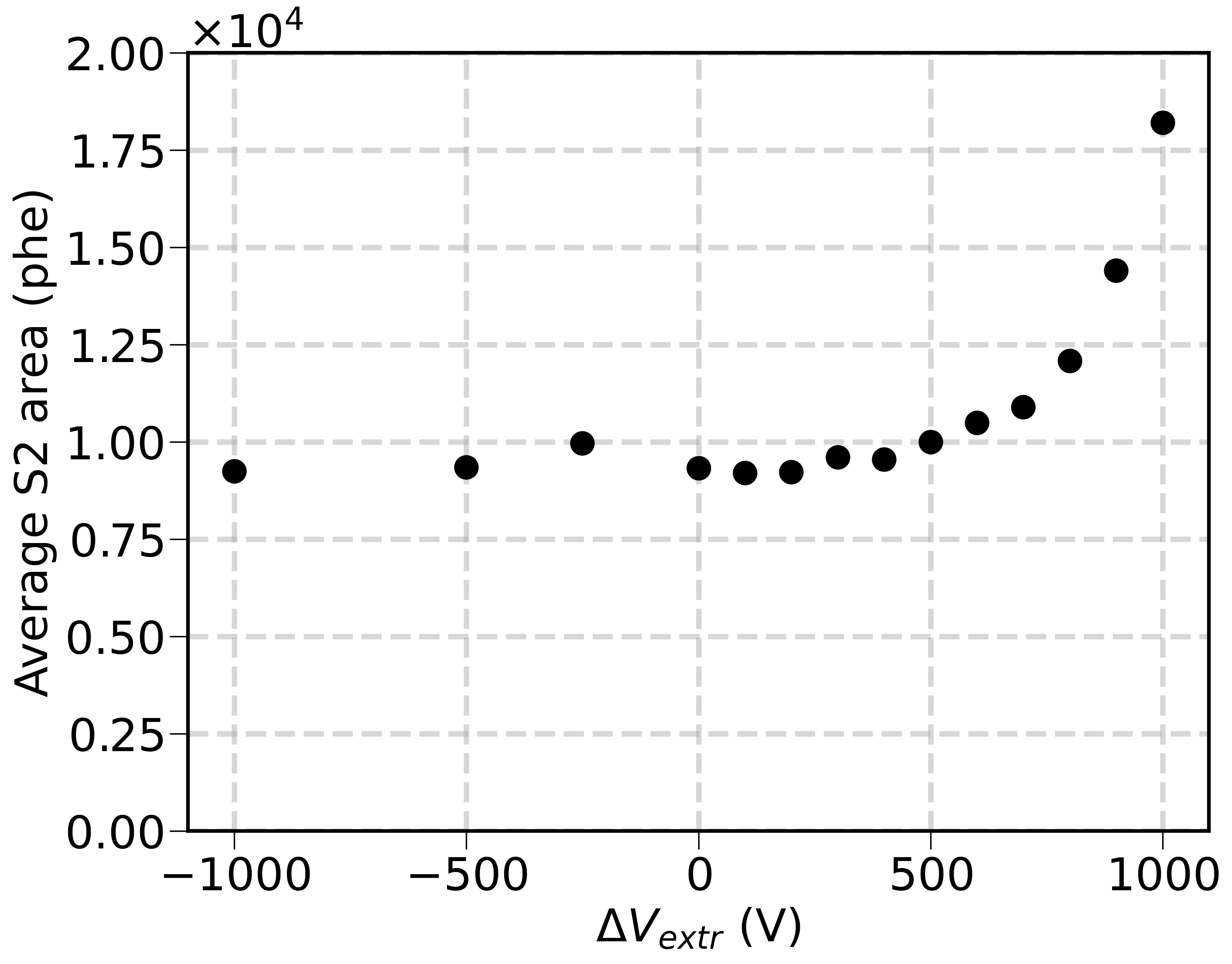} 
    \caption{Left: \stwo area as a function of \dvextr at \dvdrift$=400$~V and \dvthgem$=2500$~V. Thickness of the liquid is estimated as 6.5~mm. Right: zoomed view to the voltage range for which light is produced only in the vicinity of the holes.
    }
    \label{fig:s2area_vs_extraction}
\end{figure}

Fig.~\ref{fig:s2area_vs_extraction} shows the \stwo area as a function of \dvextr at constant values of \dvdrift$=400$~V and \dvthgem$=2500$~V. Two regions are clearly visible --- one corresponding to secondary scintillation in the THGEM holes (at low or even reversed voltage above the THGEM) and the other with a linear rise that corresponds to secondary light generation in the uniform field. The pulse area is expressed in number of photoelectrons at the PMT photocathode. The PMT calibration was done with a pulsed LED operated in a regime when only <1/20 triggers resulted in an anode signal, to ensure that the latter results from a single photoelectron.

\begin{figure}
    \vspace{5mm}
    \centering
    \includegraphics[width=7cm]{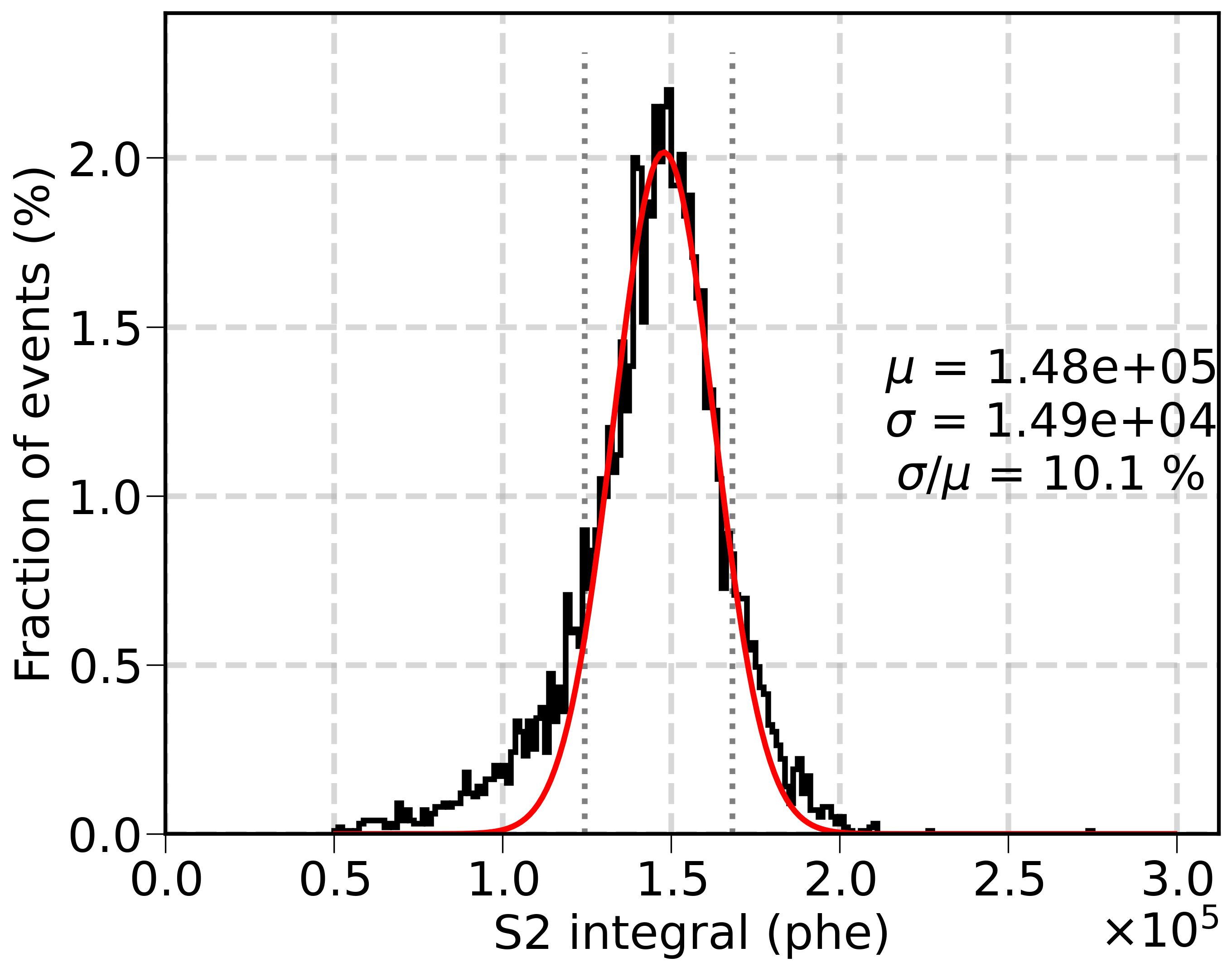} 
    \caption{\stwo spectrum of $\alpha-$particle signals for \dvdrift $= 400$~V, \dvthgem $= 2500$~V and ${\dvextr = 2500}$~V.}
    \label{fig:s2_spectrum}
\end{figure}

\begin{figure}
    \centering
    \includegraphics[width=7cm]{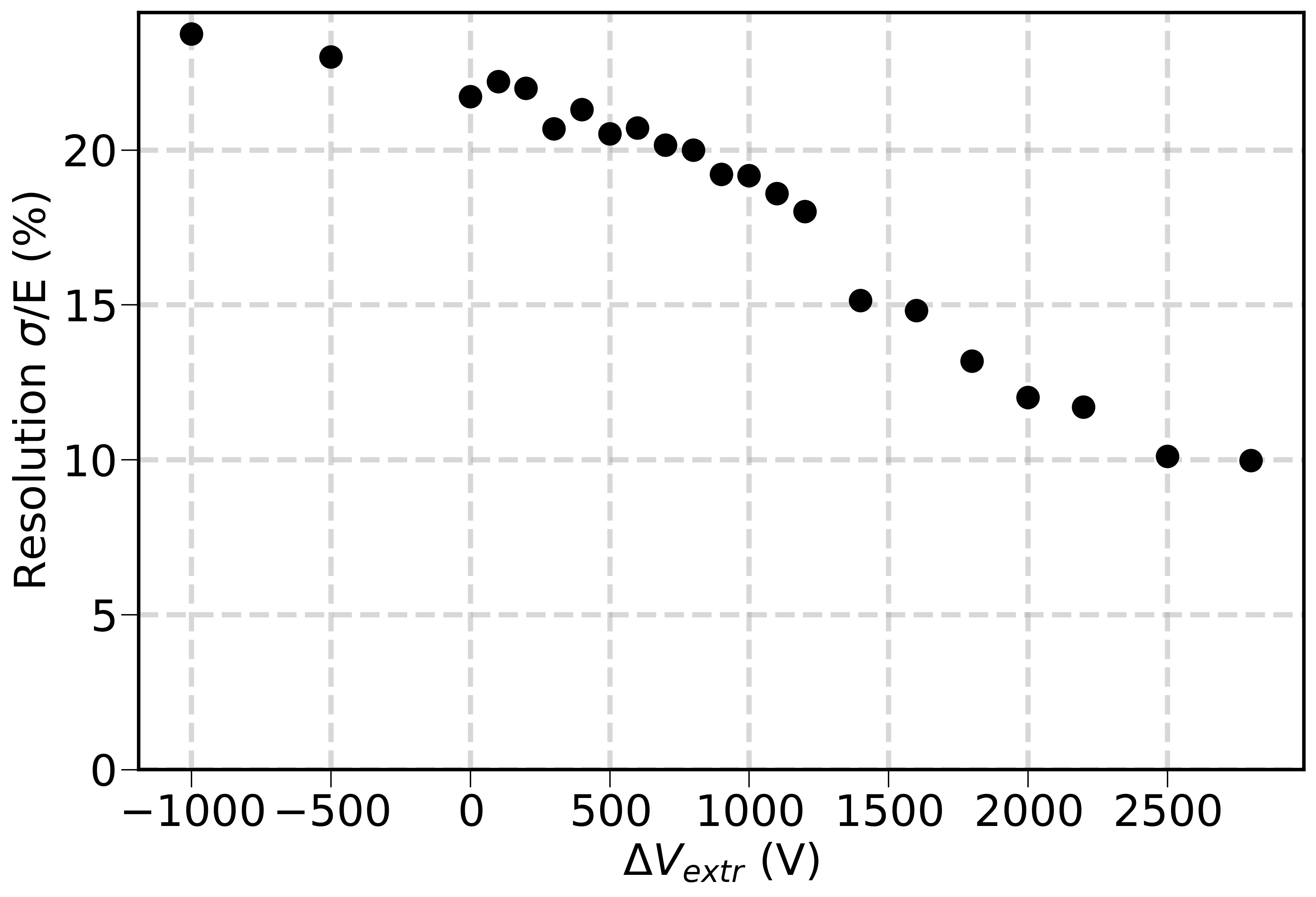} 
    \caption{Energy resolution from the \stwo spectrum as a function of \dvextr. Other voltages are fixed at \dvdrift$=~400$~V and \dvthgem$=~2500$~V.}
    \label{fig:e_resolution}
\end{figure}

An example of the \stwo spectrum is presented in Fig.\ref{fig:s2_spectrum}, while Fig.\ref{fig:e_resolution} shows the variation of the energy resolution with the extraction field. As expected, better resolution is obtained with electroluminescence in the uniform field above the THGEM.

%% file: src/discussion.tex
\section{Discussion}
\label{sec:discussion}

The above results clearly indicate that the THGEM does float on the surface of LXe, accompanying its displacement when the chamber is filled with more liquid or, oppositely, the liquid is being removed from the chamber. 

Under proper applied potentials, the electrons drifting in liquid xenon are focused into the holes of the floating THGEM and extracted into gas where they produce secondary scintillation. The observations seem to indicate that the THGEM holes are only partially filled with the liquid, although no conclusion can be drawn on the exact liquid level within the holes or the surface profile. The decrease of the light yield with an increase in the voltage across the THGEM above a certain value (Fig.~\ref{fig:s2_area_vs_vthgem}) can be explained by a deeper penetration of LXe into the holes under the electric field gradient. If this explanation is true, then LXe does not fill the hole entirely and some gas pocket exists above the liquid in the hole where secondary scintillation develops.

Using the average energy expended by a 5.486~MeV $\alpha-$particle to produce a scintillation photon in LXe, $W_s=$17.1~eV \cite{Chepel:2005}, one estimates the number of emitted photons at the $\alpha-$source to be $\approx$3$\times10^5$. Taking into account the solid angle defined by the PMT photocathode (0.063), the optical transparency of the THGEM (0.16) and that of the mesh (0.81), the PMT quantum efficiency value of $\approx$30\% for xenon wavelength~\cite{Hamamatsu}, one arrives to an yield of $\sim$730 photoelectrons per $\alpha-$particle at the PMT photocathode. 
On the other hand, the integral of the S1 signal calibrated to phe with the method described above results in a value of $\sim$450~phe per $\alpha-$particle. Such discrepancy may be due to the meniscus lensing of the liquid-to-gas interface. Although no direct information on the contact angle of liquid xenon with FR4 is available at present, the liquid-gas interface can be expected to have a concave shape within the THGEM holes: for liquid argon the angle of 48$\pm$10$^\circ$ has been measured~\cite{Tesi:2021}; for liquid xenon in contact with copper it was found to be $\approx$70$^\circ$~\cite{Baidakov:2009} (although at a much higher temperature).

The measured energy resolution is rather poor and is far from the limit defined by the photoelectron statistics. This indicates that other sources of fluctuations exist, most likely having their roots in the electron dynamics in the THGEM: focusing into the holes, passage through the hole in liquid, crossing the liquid-gas boundary, and drift in the gas to be collected to the upper THGEM electrode or extracted to the region above the THGEM. 

The THGEM tested in this work was of typical geometry widely used in gaseous detectors. The new application requires its optimization in what concerns the hole dimensions and the plate thickness, in particular. One can prospect, for example, that larger holes and a thicker plate would have an advantage of leaving more space for the gas pocket within the hole and thus leading to a higher secondary light output. From the point of view of light collection to a photon detector (presumed to be placed in gas), it can be advantageous to make conical holes with wider opening on the gas side (similarly to those discussed in \cite{Erdal:2018} for Liquid Hole Multipliers). The natural way of optimization of the THGEM geometry is its modelling using finite element method. This would require, however, a better knowledge of the wettability of materials, FR4 in the first place, by liquid xenon. 


%% file: src/conclusions.tex
\section{Conclusions}
\label{sec:conclusions}

Electron emission from LXe with a THGEM freely floating on the surface of the liquid has been observed. Also, secondary scintillation in the THGEM holes or their close vicinity from the gas side was measured as well as that in the uniform field above the THGEM. In this way, the concept of floating hole multiplier has been proven.

%% file: src/acknowledgments.tex
This work was supported by Fundação para a Ciência e Tecnologia through project CERN/FIS-INS/0026/2019. We are indebted to Mr. Y. Asher of the Weizmann Institute of Science for his technical assistance. V.C. acknowledges the personal support of the Weizmann Institute Visiting Professor Program.